# X-ray Diffraction Study of Superstructure in GdBaCo$_2$O$_{5.5}$

Yu.P. Chernenkov[1], V.P. Plakhty[1], V.I. Fedorov[1], S.N. Barilo[2], S.V. Shiryaev,[2] and G.L. Bychkov[2]

[1]*Petersburg Nuclear Physics Institute RAS, Gatchina, 188300 St.Petersburg, Russia*
[2]*Institute of Solid State and Semiconductor Physics. National Academy of sciences. Minsk 220072, Belarus*

A single crystal of GdBaCo$_2$O$_{5.47(2)}$ has been studied by means of X-ray diffraction. Appearance of superstructure reflections at $T = 341.5(7)$ K gives an evidence of continuous transition to the phase with unit cell doubled along the shortest edge $a_1$. Critical exponent for the order parameter is found to be $\beta = 0.33(1)$. The superstructure reflections are about 2-4 orders of magnitude weaker than the basic ones. Their systematic extinction indicates the crystal symmetry change from *Pmmm* to *Pmma*. The integrated intensities allow to calculate displacements of atoms from the positions in the high-temperature phase. The cobalt-ligand distances in the ordered phase are discussed in terms of the spin-state/orbital ordering of Co$^{3+}$ ions.

Since discovery of giant magnetoresistance in the oxygen-deficient layered perovskites $R$BaCo$_2$O$_{5+x}$, where $R$ is a rare earth [1], these materials attract high interest. Their orthorhombic structure at $x \approx 0.5$ with the unit cell $a_1 \approx a_p$, $a_2 \approx 2a_p$, $a_3 \approx 2a_p$, where $a_p$ is parameter of the pseudocubic perovskite cell, is usually described by the *Pmmm* space group [2–5]. One can imagine the structure as a sequence of stacking plains [CoO$_2$][BaO][CoO$_2$][$R$O$_x$] along [0,0,1], which results in two types of the cobalt environment: CoO$_5$ pyramids and CoO$_6$ octahedra. The nominal valence of cobalt at $x = 0.5$ is 3+. It is known [6] that the Co$^{3+}$ ion has a non-magnetic, or low-spin ground state (LS, $t_{2g}^6 e_g^0$) as well as two excited states, the intermediate-spin (IS, $t_{2g}^5 e_g^1$) and the high-spin (HS, $t_{2g}^4 e_g^2$) ones. The energy differences are small enough to gain the excited states by the thermal fluctuations or due to the lattice change, which results in spin state transitions [7]. A metal-insulator (MI) transition has been also found, with the transition temperature for GdBaCo$_2$O$_{5+x}$ 350 K $< T_{MI} <$ 370 K being dependent on the oxygen content of $0.4 < x < 0.47$ [1, 2 8–10]. The transition is of the first order, which is indicated by a hysteresis of 8 K in resistivity for TbBaCo$_2$O$_{5.4}$ [8].

In spite of numerous studies of these materials, neither Co$^{3+}$ spin state nor nature of the MI transition is finally established. A spin-state transition coupled with the orbital degrees of freedom is suggested to be a driving force for the MI transition. The distribution of the IS $e_g$ orbitals $(3x^2 - r^2)$ in pyramids and $(3y^2 - r^2)$ in octahedral sites on cooling has been suggested as an origin of transition on the basis of structural studies of TbBaCo$_2$O$_{5.5}$ [11]. On the other hand, it has been concluded that the transition to metallic phase in GdBaCo$_2$O$_{5.5}$ is due to excitation of the LS-state electrons into $e_g$ band of the Co HS-state in octahedra; with Co in pyramids having IS both sides of $T_{MI}$ [9]. This conclusion has been made because of octahedron expansion of about 0.012(4) Å and simultaneous pyramid shrinking. The spin-state as well as the orbital ordering among one type of coordinating polyhedra was considered in a number of publications. The neutron diffraction data for the Ho (Tb) materials have been explained by the ordering of LS and IS states in octahedra, with the same spin state in the rows along [0,0,1] alternating in the [1,0,0] direction [12]. The ordering of $e_g$ orbitals $(3z^2 - r^2)$ and $(x^2 - y^2)$ among the pyramidal sites in a chess-board like manner has been suggested in Ref. [13], with Co$^{3+}$ in the octahedral sites being considered as LS [14]. On the basis of density-functional theory calculations it has been shown that the MI transition is accompanied by a $t_{2g}$ ($xy/xz$) orbital ordering. As predicted in Ref. [15] and Rfs. therein, different spin/orbital should order among pyramidal or octahedral sites resulting in a superstructure which apparently should double the shortest edge $\mathbf{a}_1$ of the unit cell. Difference of the Co$^{3+}$ electronic structures and corresponding ionic radii is too weak effect to influence considerably intensities of the basic Bragg reflections in X-ray diffraction [16]. However the superstructure reflections, being explicitly due to atomic displacements, can give very precise data on the shape of coordinating polyhedra, which contains more information than ionic radius itself. The objective of our experiment is a search for the superstructure reflections that should allow, if any, to make some conclusions on the ordering nature.

A plate-like (~2×2×0.2 mm$^3$) single crystal of GdBaCo$_2$O$_{5+x}$ with [0,0,1] axis perpendicular to the plate was grown from the high temperature flux melt using an off-stoichiometric mixture of corresponding oxides [17]. The as-grown crystal was annealed for several days at 600 °C under 3 bars of oxygen pressure and then cooled down to room temperature. Crystal phase purity and its cation composition were checked by X-ray diffraction and X-ray fluorescent analyses. Oxygen content in the crystal that was grown and oxygenated in the same batch was determined by iodometric titration to be 5.47(2). In agreement with [10,13], the magnetization data clearly show significant anisotropy of the magnetic response. A spin-ordering transition to the phase with a spontaneous moment at $T = 267$ K is followed by transition at $T = 249$ K to a purely antiferromagnetic phase and then by second transition around $T = 150$ K to another antiferromagnetic state. The first order MI transition reveals at $T_{MI} = 365$ K in resistivity and magnetization.

A twinned structure of [1,1,0] type in the oxygenated crystals was clearly seen by Faraday image in the normal-reflection mode. The twinning is shown schematically on the reciprocal lattice inset in Fig. 1. Big and small circles correspond to the fundamental and to the superstructure reflections, respectively The indices for the doubled unit cell ($2\mathbf{a}_1, \mathbf{a}_2, \mathbf{a}_3$) are used, i.e., the superstructure points in the reciprocal lattice should have $h = 2n + 1$. However, a wide X-ray beam used for increasing the luminosity leads to low resolution. As a result, ($hkl$) and ($khl$) reflections are not distinguished because of $2a_1 \approx a_2$. Therefore only the superstructure reflections with $h$ and $k$ odd are not superimposed onto the fundamental ones. With the resolution available all the spots resulting from splitting of a reciprocal-lattice point participate in one Bragg peak, and one has to average properly over different twins when processing the data.

The X-ray beam from a conventional 2 kW tube with Mo anode was monochromated by a PG-crystal ($\lambda = 0.71$ Å). To avoid a $\lambda/2$ contamination the high voltage on the tube was kept less than threshold of the $\lambda/2$ excitation in the white spectrum. Since the superstructure reflections were expected to be very weak, rather wide beam of 0.5 mm in diameter was used. The measurements have been carried out in the transmission geometry, with the crystal always capturing the entire beam. This condition, which is necessary to obtain a quantitative set of integrated intensities, has restricted a number of independent superstructure reflections to 24. All of them as well as 11 fundamental ones have been measured at room temperature by means of the $\theta$-$2\theta$-scans. Because of poor resolution the lattice parameters $a_1$, $a_2$, $a_3$ from Ref. [2] have been used to obtain the orientation matrix. It is worth while to note that average intensity of the superstructure reflection is 2-4 orders of magnitude less than that of the fundamental ones, and is very sensitive to erroneous effects. This is why all of them have been checked by rotation around scattering vector for possible contribution of multiple scattering. To get the reasonable statistics the counting time in each point when scanning across the superstructure peaks was equal to 900 s, whereas a scan across the strongest fundamental peak took about 30 s. A typical scan is shown in Fig. 1.

Intensity temperature dependence of the superstructure peak (3,1,1) is shown in Fig. 2. The temperature of sample was changed in the range $130 \div 440$ K by the nitrogen gas flow, with the temperature stability better than 0.5 K. The thermocouple was safely glued together with the crystal so that the temperature difference between thermocouple and sample was not more than 0.5 K. Therefore one can conclude that the superstructure reflections appear below $T_{MI}$, even if there is some difference of its temperature for two crystals from the same batch. Continuous character of the transition observed gives additional argument that is not a MI transition. Intensity temperature dependence shown in Fig.2 was corrected for the critical diffuse scattering using the values above the critical point $T_C$ and assuming the ratio of the critical amplitudes $C^+/C^- = 2$ in the mean-field approximation [18]. Intensities thus corrected for a current critical point $T_{Ci}$ were fitted by the power function $I(\tau) \propto \tau^{2\beta}$, where $\tau = (T_C - T)/T_C$, and $\beta$ is critical exponent for the order parameter (atomic displacement). Then the entire procedure was repeated at another $T_{Ci}$. The final values of $T_C = 341.5(7)$ K and $\beta = 0.33(1)$ have been chosen using criteria of the residual least squares minimum. The value of $\beta$ is in better agreement with $\beta = 0.325(1)$ than with $\beta = 0.365(1)$ as obtained from the field theory in 3 dimensions for O(1) and O(3) symmetry, respectively [19]. The peak intensity follows this power law down to about 280 K, i.e. roughly to the spin-ordering transition.

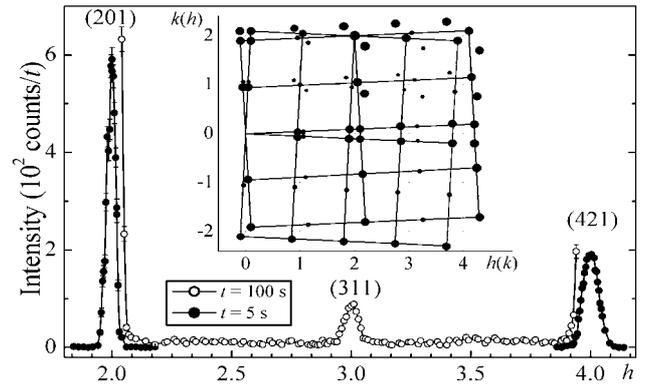

FIG. 1. A scan along ($h,h,0$) through the superstructure reflection (3,1,1). The insert shows schematically the twins for a reciprocal lattice layer $l \neq 0$.

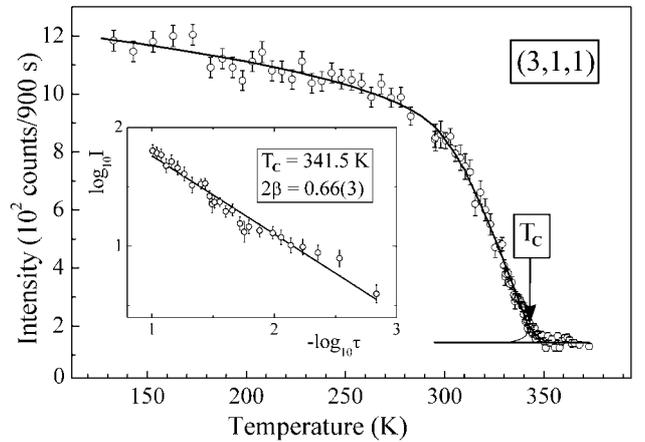

FIG. 2. Intensity temperature dependence of the (3,1,1) superstructure reflection. The insert shows The log-log plot of the intensity versus reduced temperature $\tau$ is shown in the inset.

Out of 24 reflections measured at room temperature, only 11 reflections with $l \neq 0$ have intensities non-equal to zero. (Some very weak reflections with $l = 0$ have been also detected, but their intensities strongly change when crystal is turned around the scattering vector, which evidences that they are due to multiple scattering.) This systematic extinction indicates a change of the space group. According to [20] the highest subgroup of *Pmmm* for the superstructure wave vector $\mathbf{k}_{20} = (\pi/a_1)[1,0,0]$ (in notation of [21]), which has the extinction law observed, is *Pmma* with the origin of the doubled unit cell at the point (−1/4, 0, 0). Atomic positions can be expressed through the mean coordinates $X_i, Y_i, Z_i$ in the high-temperature *Pmmm* phase and the displacements $x_i, y_i, z_i$ [22] as shown in Table I. Unlike space group *Pmmm* with equivalent positions in the doubled unit cell at $Z+z$ and $-Z-z$, one above the other, in *Pmma* they are connected by the glide plane $a$ with a difference in $X$ coordinate of 1/2. This means that the same states of $Co^{3+}$ (spin or orbital) are not arranged in the rows along [0,0,1], but are ordered in a chess-board like manner in both octahedral and pyramidal planes.

TABLE I
Distribution of atoms over positions of space group *Pmma*.

| Site | Coordinates | Atoms |
|------|-------------|-------|
| 2e | 1/4, 0, Z+z | 2Co1P, 2Co2P, 2O1 |
| 2f | 1/4, 1/2, Z+z | 2Co1O, 2Co2O, 2O2, 2O3 |
| 4g | 0, Y+y, 0 | 4Ba |
| 4h | 0, Y+y, 1/2 | 4Gd |
| 4i | X+x, 0, Z+z | 4O4 |
| 4j | X+x, 1/2, Z+z | 4O5 |
| 4k | 1/4, Y+y, Z+z | 4O61, 4O62 |

The scale factor for calculation of atomic displacements has been obtained from the basic Bragg intensities using for $X_i, Y_i, Z_i$ the data [4], but there is no significant difference in the scale factor obtained with the set of data [2]. All necessary corrections of the intensity were applied in calculations: absorption, irradiated sample volume, polarization, and Lorentz-factor. The extinction correction was insignificant. The superstructure intensities were corrected in the same way and used in refinement of the atomic displacements at fixed value of the scale factor. Contribution of the twins (Fig. 1) was assumed to be equal in calculation. Owing to small values of displacements $x_i, y_i, z_i$, one can express intensity of a superstructure reflection with the Miller indices $h, k, l$ through the structure factor developed into series to the first order in displacements from the mean atomic positions as

$$I(q) \propto \{\sum 2\pi f(q)(hx_i + ky_i + lz_i)\cos[2\pi(hX_i + kY_i + lZ_i)]\}^2, \quad (1)$$

where $q = \sin\theta/\lambda$, $f(q)$ is the atomic form-factor, and the sum is over all the atoms of doubled unit cell, which participate in the structure factor. For instance, the atoms of Ba and Gd enter the intensities of reflections ($hkl$) with $h = 2n$, which have not been measured. Since $X(O4) = X(O5) = 0$, $z$−displacements of these atoms do not participate in the intensity while those of O1, O2 cannot be distinguished and should be refined as one parameter. As a result there are 12 independent displacements, as follows from Table I. Having intensities of 11 superstructure reflections, we are obliged to make some reasonable assumptions. We have assumed at the beginning that $|z(Co1P)| \approx |z(Co2P)|$ and $|z(Co1O)| \approx |z(Co2O)|$, which finally occur to be zeros in the limits of the standard deviations, and we have left with 7 variables. The results of refinement are shown in Fig. 3, and the agreement between calculated and observed intensities in Fig. 4. The least squares residual is $\chi^2 = 2.5$.

Some conclusions on the ordering models mentioned above can be made even without numerical results. The ordering of LS and HS states in octahedra [12] is described by space group *Pmmm* while the orbital ordering in pyramids [13] results in vanishing of some superstructure reflections. Therefore these two models can be ruled out. Another important conclusion that can be made on the basis of qualitative results concerns the nature of spontaneous moment. Until now, the unit cell doubling was considered to be a result of antiferromagnetic ordering with the wave vector $\mathbf{k}_{20} = (\pi/a_1)[1,0,0]$. In the unit cell doubled because of atomic displacements the wave vector of magnetic structure is equal to zero, which is an obligatory condition for appearance of a spontaneous moment.

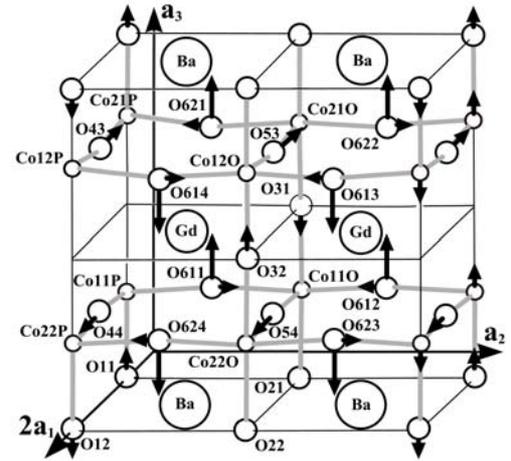

FIG. 3. Displacements of atoms in the *Pmma* phase from their mean positions in the *Pmmm* phase. The origin is at the point (−1/4, 0, 0). The arrow lengths are approximately proportional to the displacements: $|z(O1+O(2)| = 0.011(2)$ Å; $|z(O3)| = 0.008(1)$ Å; $|x(O4)| = 0.0304(5)$ Å; $|x(O5)| = 0.0437(7)$ Å; $|y(O61)| = 0.0011(5)$ Å; $|z(O61)| = 0.031(2)$ Å; $|y(O62)| = 0.0043(5)$ Å; $|z(O62)| = 0.036(4)$ Å.

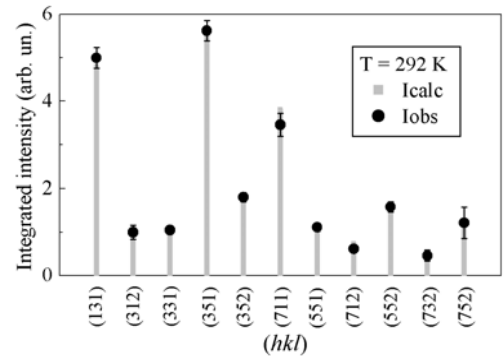

FIG. 4. Comparison of calculated and observed intensities of the superstructure reflections.

One cane make further conclusions from distances between the $Co^{3+}$ ions and the $O^{2-}$ ligands (Table II), which determine the shapes of octahedra and pyramids. Some uncertainties arise since only the sum $z(O1) + z(O2)$ instead of the individual displacements $z(O1)$ and $z(O2)$ is available, but these uncertainties do not exceed 0.01 Å as seen from the caption to Fig.3. The situation is clearer for the octahedral sites. Four Co−O distances in the **xy** plane of the octahrdra Co11O and Co12O are almost equal while for Co21O and Co22O the distance along **x** is about 0.1 Å shorter than along **y**. Taking into account that the distance along **z** is also ~0.1 Å shorter in both cases, one can suggest $e_g$ orbitals $(x^2 - y^2)$ and $(3y^2 - r^2)$ for the first and the second pair, respectively as shown in Fig. 5. For Co11P, Co12P $e_g$ orbital $(3x^2 - r^2)$ seems to be obvious. The situation with Co21P and Co22P electron configurations is not that clear. The distances to all five ligands are roughly the same, somewhat in between the shortest and the longest octahedral distances. As seen in Fig. 2 of Ref. [4], interatomic distances in pyramids are in average shorter than in octahedra. Taking this observation into account, we assume that $e_g$ orbitals in Co21P and Co22P are $(x^2 - y^2) + (3z^2 - r^2)$, i.e., these ions are probably in HS state as has been observed for pyramids in $Sr_2CoO_3Cl$ [23]. However, a pure $t_{2g}^6$ LS state cannot be excluded. The assumption [3] that orbital ordering exists only in the pyramidal sites is not consistent with our results. According to [4], the octahedral $Co^{3+}$ ions should be in LS state below the MI transition, while in our case they are in IS state. From our data we can say nothing as to the $t_{2g}$ ($xy/xz$) orbital ordering [14].

TABLE II
Distances (Å) between $Co^{3+}$ and ligands with numerals in Fig. 3.

| Octahedra | O51, O52 | O61, O62 | O2 | O3 |
|---|---|---|---|---|
| Co11O | 1.989(7) | 2.023(6) | 1.900 | 1.859(2) |
| Co21O | 1.902(7) | 2.018(7) | 1.900 | 1.875(2) |
| Pyramids | O41, O42 | O61, O62 | O1 | – |
| Co11P | 2.016(6) | 1.943(7) | 1.919 | – |
| Co21P | 1.956(6) | 1.928(7) | 1.941 | – |

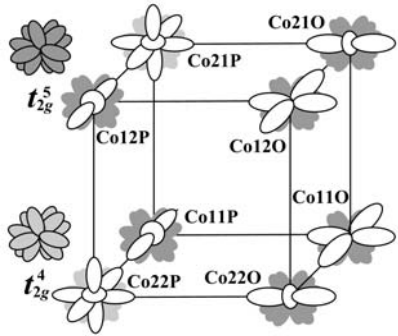

FIG. 5. Spin state/orbital ordering among octahedral (Co11O – Co22O) and pyramidal (Co11P – Co22P) sites. The orbitals $e_g$ are described on the background of $t_{2g}^4$ and $t_{2g}^5$ ones.

In conclusion, a second order transition at $T_C = 341.5(7)$ K is discovered to a phase with the symmetry *Pmma*. The $\mathbf{a}_1$ edge of the high temperature unit cell (*Pmmm*) is doubled. The atomic displacements are obtained from intensities of the superstructure reflections. The cobalt-oxygen distances in the pairs of non-equivalent pyramids and octahedra allow making some conclusions on the orbital ordering of IS $Co^{3+}$ ions among octahedral sites. Apparently there is HS-IS ordering in pyramidal sites. Similar results, which will be published elsewhere, are obtained also for $DyBaCo_2O_{5.5}$.

This work was carried out in framework of the INTAS project (Grant N 01-0278). We are grateful for a partial support by the Russian Foundations for Fundamental Researches (Project N° 02-02-16981), by the grant SS-1671.20032, and by the NATO grant PST CLG 979369.

[1] C. Martin *et al*., Appl. Phys. Let. **71**, 1421 (1997).
[2] M. Respaud, *et al*., Phys. Rev. B **64**, 214401 (2001).
[3] S. Roy, M. Khan, *et al*., Phys. Rev. B **65**, 064437 (2002).
[4] C. Frontera *et al*., Phys. Rev. B **65**, R180405 (2002).
[5] C. Frontera *et al*., J. Solid State Chem. **171,** 349 (2003).
[6] M.A. Korotin *et al*., Phys. Rev. B **54**, 5309 (1996).
[7] K. Asai, P. Gehring *et al*., Phys. Rev. B **40**, 10982 (1989).
[8] A. Maignan *et al*., J. Solid State Chem. **142,** 247 (1999).
[9] C. Frontera *et al*., Phys. Rev. B **65**, R180405 (2002).
[10] D.D. Khalyavin *et al*., Phys. Rev. B **64**, 214421 (2003).
[11] Y. Moritomo *et al*., Phys. Rev. B **61**, R13325 (2000).
[12] F. Fauth, E. Suard *et al*., Phys. Rev. B **66**, 184421 (2002).
[13] A.A. Taskin, A.N. Lavrov, and Y. Ando, Phys. Rev. Lett. **90**, 227201 (2003).
[14] Hua Wu, Phys, Rev B **64**, 092413 (2001).
[15] D.I. Khomskii, U. Low, Phys. Rev. B **69**, 184401(2004)
[16] P.G. Radelli and S.-W. Cheong, Phys. Rev. B **66**, 094408 (2002)
[17] D.D. Khalyavin *et al*., Phys. Rev. B **67**, 214421 (2003).
[18] E. Brezin, J.C. Le Guillou, and J. Zinn-Justin, in *Phase Transitions and Critical Phenomena*, Vol. 6, edited by C. Domb and M. S. Green (Academic, London, 1976), Vol.VI.
[19] J.C. Le Guillou and J. Zinn-Justin, Phys. Rev. Lett. **39**, 95 (1977)
[20] V.E. Naish, S.B.Petrov and V.N.Syromyatnikov. *Subgroups of the space groups – II. Subgroups with changing unit cell*. Preprint No. 486-77Dep. (Institute of Metal Physics, RAS, 1977). In Russian.
[21] O. V. Kovalev, *Irreducible representations of the space groups* (Gordon and Breach, 1965).
[22] *International Tables for Crystallography*. Vol. A, edited by Th. Hahn (D. Reidel Publishing Company, Dordrecht – Boston, 1983).
[23] Z. Hu *et al*., Phys. Rev. Lett. **92**, 207402 (2004).